\documentclass[12pt]{iopart}

\usepackage{graphicx}
\usepackage{amsfonts}

\usepackage{epsfig}
\usepackage{graphicx}
\usepackage{dcolumn}
\usepackage{bm}
\usepackage{times}
\usepackage{xcolor}
\usepackage{colortbl,booktabs}
\usepackage{makecell}
\usepackage{amstext}
\usepackage{amssymb} 
\usepackage{float}
\usepackage{verbatim}

\begin{document}

\title[Community detection on complex networks ]{Community detection on complex networks based on a new centrality indicator and a new modularity function
}    

\author{Junfang Zhu$^{1,2}$, Xuezao Ren$^{1}$, Peijie Ma$^{1}$ and Kun Gao$^{1}$ } 

\address{$^1$School of Science, Southwest University of Science and Technology, Mianyang 621010, People's Republic of China}

\address{$^2$Big Data Research Center, University of Electronic Science and Technology of China, Chengdu 611731, People's Republic of China}

\ead{renxuezao@aliyun.com}

\begin{abstract}

Community detection is a significant and challenging task in network research. Nowadays, plenty of attention has been focused on local methods of community detection. Among them, community detection with a greedy algorithm typically starts from the identification of local essential nodes called central nodes of the network; communities expand later from these central nodes by optimizing a modularity function. In this paper, we propose a new central node indicator and a new modularity function. Our central node indicator, which we call local centrality indicator ($LCI$), is as efficient as the well-known global maximal degree indicator and local maximal degree indicator; on certain special network structure, $LCI$ performs even better. On the other hand, our modularity function $F2$ overcomes certain disadvantages—such as the resolution limit problem—of the modularity functions raised in previous literature. Combined with a greedy algorithm, $LCI$ and $F2$ enable us to identify the right community structures for both the real-world networks and the simulated benchmark network. Evaluation based on the normalized mutual information ($NMI$) suggests that our community detection method with a greedy algorithm based on $LCI$ and $F2$ performs superior to many other methods. Therefore, the method we proposed in this paper is potentially noteworthy.

\end{abstract}

\maketitle


\section{Introduction}
Community is a basic feature of complex network on mesoscopic scale.\cite{Mucha2010}. It has been found in different types of complex networks, including  social networks\cite{Zachary1977,Girvan2002,Lusseau2004}, biological networks with function modularity\cite{Jonsson2006}, technological networks\cite{Newman2004a} and so on. A community is a group of nodes in a network, which have denser connections (edges) among themselves than with other nodes outside the group\cite{Girvan2002}. Community detection is of significant importance for network science\cite{Fortunato2010}. However, a strict definition for community structure is still to be recognized. \cite{Fortunato2010,Ciglan2013}. And it has been always a huge challenge to detect communities within a large network.

In past decades, many methods of community detection have been proposed, such as graph partitioning\cite{Shi2011}, spectral clustering\cite{Donetti2004,Capocci2005,Jin2015}, statistic inference\cite{Newman2007}, dynamic methods\cite{Zhou2003,Hu2008}, and integer programming model\cite{Srinivas2019}.
Especially, one class of methods that employs greedy algorithms to optimize a quality function has attracted much attention\cite{Newman2004b,Clauset2004}. Such methods usually adopt a the nearest search technique to expand communities\cite{Saha2016}. Two factors are vital for the accuracy of community detection with a greedy optimization algorithm. One is the selection of "central nodes" from which community expansion starts. Starting from "marginal nodes" instead of "central nodes," the result could be very different. Therefore, it is essential to discover actual "central nodes" by which local communities can be identified correctly\cite{Chen2013}.
It has been realized that nodes with global maximal degrees may not all be central nodes, especially when such nodes are adjacent to each other, or all involved in a same community\cite{Chen2013}.Therefore, Chen et al. selected nodes with local maximal degrees which locate dispersedly in the network as candidates for central nodes\cite{Chen2013}.
However, a reliable way of central node selection is still an open problem.

The other vital factor for community detection is the objective function for the greedy optimization process, in which modularity functions measuring the "strength" of community structure are often used. One famous modularity function is the modularity $Q$ as defined by Newman\cite{Newman2004a} (see section 2 for its definition). It has been applied to many community detection algorithms, including the Newman fast algorithm\cite{Newman2004b}, EO algorithm\cite{Duch2005}, and greedy optimal algorithm\cite{Blondel2008}.
However, the calculation of $Q$ requires certain global information of the whole network, which seriously increases the computational burden, especially for large networks. Besides, $Q$ has been realized to have a resolution problem\cite{Santo2007}. In past years, a number of local modularity functions have been proposed, such as $R$\cite{Aason2005}, $M$\cite{Luo2008} and $F$\cite{Lancichinetti2009} and the modularity density $D$\cite{Li2008}. These modularity functions on the one hand overcome certain drawbacks of modularity $Q$, but on the other hand also have their own problems. For example, when an isolated subgraph within the network emerges as a community, or the whole network constitutes a single community, $R$ is lack of definition, and $M$ approaches infinite. For certain communities under optimal partition, the local modularity density $D$ becomes negative\cite{Li2015}. The performances of these existing modularity functions are far from satisfactory; new functions of modularity are to be discovered.

In this paper, we propose a new method of community detection. Based on local information of network, we suggest a new indicator for central nodes, as well as a new modularity function of communities. Our greedy algorithm of community expansion allots residual nodes to communities detected by the new indicator for central nodes. Such an algorithm is applicable to real-world networks, and has been evaluated by normalized mutual information ($NMI$)\cite{Kuncheva2004} on benchmark networks through computer simulation. 

\section{Related work}

In this section, we briefly introduce some community detection methods raised in previous literature. We mainly introduce the central nodes indicators and modularity functions used in these methods. In latter part of this paper, we will compare the performances of these methods to the performance of our method proposed in this paper. 

\subsection{Central nodes indicators}

In a heterogeneous network, the "importance" of nodes is not even; some nodes are obviously more important than other nodes\cite{lu2016}. Important nodes are called "central nodes" for community detection. Starting from central nodes rather than other less important nodes, a community expansion algorithm usually ends up with better outputs. It has been observed that nodes with higher degrees are in most cases more important than nodes with lower degrees for community detection. In previous literature\cite{Chen2013}, nodes with either global maximal degrees or local maximal degrees are selected as central nodes. 

The global maximal degree indicator takes nodes of top $k$ highest degrees as central nodes. However the selection of the value of k is often arbitrary. On the other hand, the global maximal degree indicator considers only the degree of a single node, which is not always sufficient. From the perspective of statistics\cite{Blondel2008b}, nodes with global maximal degrees are not always central nodes. 

In contrast to the global maximal degree indicator, the local maximal degree indicator designates a node as a central node only if it has the highest degree among all its neighbors. Central nodes designated by the local maximal degree indicator usually locate dispersedly in the network, except when two or more adjacent nodes have equal degrees and are all designated as central nodes. Studies have revealed that the local maximal degree indicator usually performs better than the global maximal degree indicator\cite{Chen2013}.  

\subsection{Modularity functions}

Modularity function measures the strength of the community structure, and is widely used as quality functions in community detection algorithms. The most popular modularity function is the Newman's modularity $Q$ \cite{Newman2004a} which is formulated as,

\begin{equation}
Q =\sum\limits_{i = 1}^m {[\frac{l_{in}(C_i)}{L}- ({\frac{d(C_i)}{2L}})^2] }
\end{equation}
Here ${l_{in}(C_i)}$ is the number of edges among nodes within community $C_i$. $d(C_i)$ is the sum of degrees of  all nodes in $C_i$. $L$ is total number of edges within the whole network. $Q$ has been proved to be effective in previous literature\cite{Newman2004a}. However, it has also been revealed that Q has some serious shortcomings\cite{Santo2007,Arenas2007,Andrea2011,Chen2013b,Chen2014,Dzamic2020}. For example, there exists a "resolution limit problem" in $Q$\cite{Santo2007,Arenas2007,Andrea2011,Dzamic2020}: when the size of a community is below a certain threshold, it can't be detected by $Q$. This threshold does not depend on particular network structure, but results only from the comparison between the number of links of interconnected communities and the total number of links of the network\cite{Santo2007}. In some cases, maximizing $Q$ tends to split a large community into smaller ones \cite{Chen2013b,Chen2014,Dzamic2020}. And some random networks, which by definition have no apparent community structure, may have unreasonably large values of $Q$\cite{Guimera2004,Reichardt2006}. All these shortcomings reflect that $Q$ is far from perfect, which urges researchers to find better functions for modularity.

The calculation of modularity $Q$ requires global information of the network. For the large networks, a modularity function based on local information has higher efficiency. For this Clauset defined a local modularity function $R$\cite{Aason2005} which directly measures the sharpness of the boundary of a local community. For a local community $D$, nodes within it are split into two subsets, $C$ and $B$. Subset $C$ is consist of "core" nodes that have connections only within $D$, and subset $B$ is consist of "boundary" nodes that have at least one connection to nodes outside $D$. Clauset defined the local modularity $R$ as

\begin{equation}
R =\sum\limits_{i = 1}^m {\frac{B_{in}(C_i)}{B_{in}(C_i)+B_{out}(C_i)}}
\end{equation}
Here $B_{in}$ is the number of edges that connect boundary nodes with nodes in $D$, $B_{out}$ is the number of edges that connect boundary nodes with nodes out of $D$. 

Besides $Q$ and $R$, two other modularity functions are also noticeable. Luo et al. \cite{Luo2008} defined a modularity measure $M$ which is defined as the ratio of the number of internal edges to the number of external edges of a community.

\begin{equation}
M =\sum\limits_{i = 1}^m {\frac{E_{in}(C_i)}{E_{out}(C_i)}}
\end{equation}

In which $E_{in}(C_i)$ and $E_{out}(C_i)$ are the number of internal edges and the number of external edges respectively.

And Lancichinetti et al.\cite{Lancichinetti2009} defined a modularity $F$ by comparing the "in-degree" of a community to $\alpha$ power of the "total-degree".

\begin{equation}
F = \sum\limits_{i = 1}^m {\frac{{d{_{in}(C_i)}}}{{{{(d_{in}(C_i) + d_{out}(C_i))}^\alpha}}}}
\end{equation}
Here $d_{in}(C_i)$ denotes the in-degree of community $C_i$, i.e., twice the number of edges within the community, and $d_{out}(C_i)$ denotes the out-degree of $C_i$ which is the number of edges connecting nodes in the community with nodes out of the community. The total-degree is the sum of the in-degree and the out-degree. As in literature\cite{Lancichinetti2009}, $\alpha$ is usually set to 1.

\section{Methods}

\subsection{A new local centrality indicator}	

To extract central nodes of communities, for each node $i$ in the network, we define a local centrality indicator (LCI) as following:

\begin{equation}
LCI_i = \frac{{{k_i} - \frac{1}{{{k_i}}}\sum\limits_{j \in {\Gamma _i}} {{k_j}} }}{{{k_i} + \frac{1}{{{k_i}}}\sum\limits_{j \in {\Gamma _i}} {{k_j}} }}.
\end{equation}

where $k_i$ and $k_j$ correspondingly represent the degrees of node $i$ and node $j$, in which node $j$ is a neighbor of node $i$ in the network. $\Gamma_i$ represents the set of all neighbors of node $i$. $LCI$ reflects the centrality of each node locally in the network relative to its neighbors. Obviously, the value of $LCI$ is between -1 and 1, and larger $LCI$ indicates higher local centrality. Without loss of generality, in this paper we take nodes with $LCI_i \geq 0$ as central nodes for community detection; other nodes are all non-central nodes.

\subsection{A new local modularity function}	

We propose a new local modularity function $F2$ for community detection. For community $C_i$ in a network, the $F2$ of $C_i$ is defined as 

\begin{equation}
F2(C_i) = \frac{[d_{in}(C_i)]^2}{[d_{in}(C_i) + d_{out}(C_i)]^2} 
\end{equation}
where  $d_{in}(C_i)$ and $d_{out}(C_i)$ are the in-degree and out-degree of community $C_i$ which are identically defined as in modularity $F$. The modularity $F2$ for the whole network is defined as the sum of $F2$ of all communities within the network.

\begin{equation}
F2 = \sum\limits_{i = 1}^m{F2(C_i)}
\end{equation}

This new modularity $F2$ will be used as an optimal function in our greedy algorithm for community detection.

\subsection{Community detection with a greedy algorithm based on $LCI$ and $F2$}	

Based on our new centrality indicator $LCI$ and new modularity function $F2$, we detect communities within a complex network with a greedy algorithm. Detailed procedures are described below.

First of all, we extract all central nodes by $LCI_i \geq 0$ in the network. Communities will then expand from these central nodes. 

Then we start the community expansion procedure. Initially, we randomly choose one from the central nodes. We take this node as a seed of a community, say $C_1$, and start expanding it. At each step of the expansion of $C_1$, we search in its unassigned neighbors, i.e., nodes that are directly connected to $C_1$ and haven't been assigned to any community. Our goal is to find one node among all these unassigned neighbors that maximizes the value of $F2(C_1)$, if we add this node to $C_1$. In case the same maximal $F2(C_1)$ can be obtained by adding different neighbors to $C_1$, we randomly choose one from these neighbors and add it to $C_1$. We repeat this step, until community $C_1$ stops expanding when any possible expansion of $C_1$ decreases its $F2$. Then we move on to the next community by randomly choosing another central node, which hasn't been assigned to any community, and expanding it exactly the same way as the expansion of $C_1$. We repeat this process, until all communities stop expanding, and there is no central nodes left unassigned in the network. It should be noted that in the above procedure, the order of choosing central nodes is random so that the order of expansions of different communities is random too. In principle, different orders may have different outputs. However, to our investigation, although the number of possible arrangements of orders is huge, these different arrangements only result in finite numbers of possible outputs. In practice, we usually implement the community expansion procedure for a number of times, and take the output with the highest $F2$ as the final output. 

Once the community expansion procedure is done, most nodes in the network should have been assigned to different communities. However, there may still exist a very small number of non-central nodes that remain unassigned. For each of these unassigned nodes, we search in its neighbors and find out the one with the highest $LCI$. We arbitrarily merge this unassigned node to the same community as its neighbor with the highest $LCI$. It should be noted that this operation may decrease the final $F2$ of the network, however it is reasonable because community structure with exactly the highest value of modularity function might be an overfit, as discussed in\cite{Zhang2014}.

\section{Results}

\subsection{Performance of the local centrality indicator ($LCI$)}

A noticeable advantage of the local centrality indicator ($LCI$) is, it does not miss local central nodes with low degrees. In contrast, the global maximal degree and local maximal degree indicators would more or less have the problem.  Figure 1 shows an example: the network is constructed by two complete graphs connected through one edge only. Obviously each complete graph constitutes a community, and nodes $1$ and $2$ are central nodes. However, a global maximal degree indicator which extracts the top $\emph{k}$ highest degree nodes as central nodes, tends to equate node $2$, which is a central node, with non-central nodes $3, 4$ and $5$.  As shown in figure 1(a), when $\emph{k=1}$, node $2$ will be classified as a non-central node; when $\emph{k}$=$2$-$5$, not only node $2$ but also nodes $3$-$5$, are identified as central nodes. In both cases, the global maximal degree indicator fails to recognize the difference between the centrality of node $2$ and the centrality of nodes $3$-$5$. On the other hand, when two central nodes of different degrees are connected to each other, a local maximal degree indicator would fail to identify the one with lower degree as a central node, as node $2$ in figure 1(b). In contrast, our $LCI$ correctly identifies node $1$ and $2$ as central nodes. The local centrality of a node can be well measured by $LCI$ in figure 1(c).

\begin{figure}[H]
\setlength{\abovecaptionskip}{0.cm}
\setlength{\belowcaptionskip}{-0.cm}
\centering
	\includegraphics[width=1\textwidth]{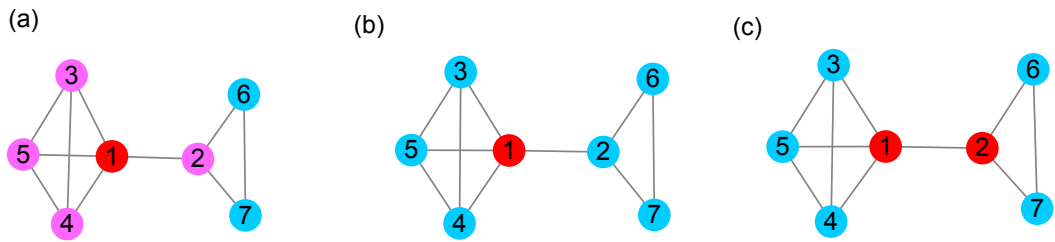}
    \caption{Performances of different centrality indicators. The network shown in the figure is constructed by two complete networks connected by one edge only. nodes $1$ and $2$ are central nodes of the network. (a) Indicated by the global maximal degree indicator, the red node (node $1$) is the top $1$ highest degree node, purple nodes (nodes $2$-$5$) are top $2$-$5$ highest degree nodes, and blue nodes are non-central nodes. (b) Indicated by the local maximal degree indicator, only node $1$ is a central node, other nodes are all non-central. (c) Indicated by our local centrality indicator ($LCI$), nodes $1$ and $2$ are central nodes. Obviously $LCI$ most correctly identifies central nodes among all three indicators. }
\label{fig_IIlustration}
\end{figure}

\subsection{Performance of the new modularity function $F2$}

Modularity functions are customarily used as optimal functions for community detection. Comparing to previous modularity functions, such as $Q$, $M$ and $F$, our modularity function $F2$ has certain advantages. It can be theoretically proven that a random network or a complete graph can't be divided into any two parts (See appendix). Unlike $Q$ and $M$, which both have a resolution limit problem\cite{Santo2007,Arenas2007,Andrea2011,Dzamic2020}, our $F2$ identifies small cliques well. In figure 2(a), we constructed a network with a series of $p$-cliques connected to a ring with single edges; each $p$-clique is a complete graph containing $p$ nodes and $p(p-1)/2$ edges. Obviously, such a network has a clear community structure that each clique corresponds to a community. For such a network, $F2$ identifies each $p$-clique as an individual community.However, $Q$ and $M$ tend to merge adjacent cliques into a bigger community, as shown by the dashed ovals in figure 2(a). In previous literature, this is called a "resolution limit problem" of $Q$ and $M$\cite{Santo2007,Arenas2007,Andrea2011,Dzamic2020}. In contrast, $F2$ does not have the resolution limit problem; it always identifies single cliques rather than merged ones as communities; a theoretical proof can be found in Appendix A.

When the network is constructed by cliques of different sizes, $Q$ and $M$ still have a resolution limit problem on smaller cliques. As in figure 2(b), the network is constructed by two $6$-cliques and two $3$-cliques. In this case, $Q$ and $M$ can identify the $6$-cliques as individual communities, but still tend to merge the $3$-cliques into a bigger community. In contrast, $F2$ still identifies all four cliques as individual communities, regardless of their sizes. More theoretical proofs can be found in Appendix A.

On the other hand, when between-clique edges increase so that different cliques tend to be well connected, $F2$ can avoid splitting a well-connected community into smaller ones. Figure 2(c) shows such an example: each $4$-clique has 6 inner edges but up to 10 between-clique edges; such a network has been recognized as a well-connected network in previous literature\cite{Chen2013b}.In this case, $Q$ and $F$ still tend to split the whole network into two communities, each consisting of a $4$-clique, while $F2$ merges these two $4$-cliques and identifies the whole network as a community.

\begin{figure}[H]
\setlength{\abovecaptionskip}{0.cm}
\setlength{\belowcaptionskip}{-0.cm}
\centering
	\includegraphics[width=1\textwidth]{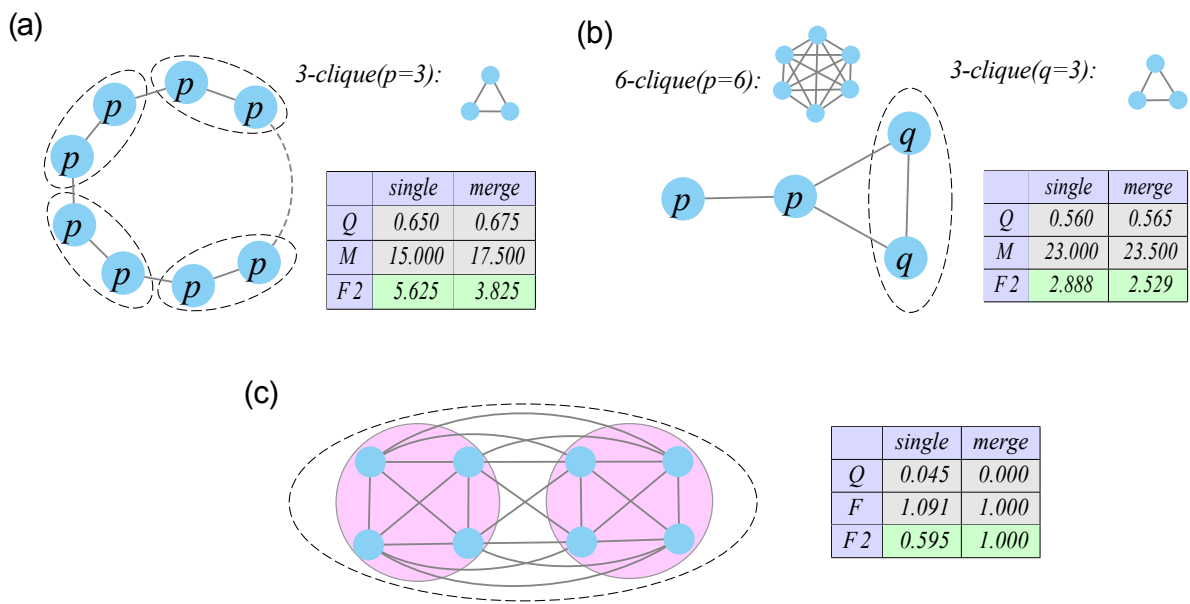}
    \caption{Performances of different modularity functions in community detection on three representative networks. (a) Ten $3$-cliques connected to a ring through single edges; each $3$-clique is a complete graph of 3 nodes and 3 edges\cite{Santo2007}. On this network, two candidate structures of community arise: one is to identify each $3$-clique as a single community, and the other is to merge each pair of adjacent cliques into one community; values of modularity functions $Q$, $M$ and $F2$ for these two structures are listed as "single" and "merge" respectively in the inserted table. (b) A network consist of two $6$-cliques and two $3$-cliques\cite{Santo2007}. "Single" in the inserted table refers to identifying each clique as an individual community, while "merge" refers to merging the two $3$-cliques into one community. (c) A well-connected network  raised in reference\cite{Chen2013b}, for which "single" splits the network into two communities, while "merge" identifies the whole network as one community.}
\label{fig_IIlustration}
\end{figure}

\subsection{Community detections in real-world networks with a greedy algorithm employing $LCI$ and $F2$}

Three real-world networks are customarily  used to estimate the performance of a community detection method: the Zachary's karate club network\cite{Zachary1977}, the dolphin network\cite{Lusseau2004}, and the college football network\cite{Girvan2002}; parameters of these networks are all listed in table 1. In this section, we will examine on these real-world networks the performance of our greedy algorithm for community detection based on $LCI$ and $F2$.

\begin{table}  
\centering  
\caption{Structure parameters of three real-world networks: karate, dolphins and football. For each network, parameters $N$, $L$ and $m$ represent the numbers of nodes, edges and communities respectively, and $\langle k\rangle$ represents the average degree of all nodes in the network. $NMI$ stands for the normalized mutual information between communities detected by our method and the reference communities proposed in previous literature\cite{Zachary1977,Girvan2002,Lusseau2004}.}  
\begin{tabular}  
{>{\columncolor{white}}rccccc}  
\hline
\rowcolor[gray]{0.9}   Networks &$N$ &$L$   &$m$ &$\langle k \rangle$ & $NMI$ \\  
\hline
karate   &34   &78  &2  &4.5882 &1 \\  
dolphins   &62 &159 &2   &5.129 &0.8904 \\   
football   &115 &613    &12&10.6609 &0.9429 \\  
\hline
\end{tabular}  
\end{table}  

For the karate network, among all $34$ nodes five nodes have positive $LCI$ values, which are nodes $34, 1, 33, 2$ and $3$, with $LCI$ values $0.6328, 0.5754, 0.4049, 0.2180$ and $0.2048$ respectively. These nodes are identified as central nodes of the network. Then we implement a greedy algorithm to maximize the $F2$ of the network, and obtain two communities:  $C_1=\{1,2,3,4,8,14,18,20,22\}$, $C_2=\{9,10,15,16,19,21,23,24,27,28,30,31,33,34\}$ in figure 3. And there are nine nodes (grey nodes in figure 3) left unassigned to either $C_1$ or $C_2$. Among them, nodes $5, 6, 7$ and $11$ are neighbors of node $1$; these nodes are later assigned to community $C_1$. Samely, nodes $29$ and $32$ are neighbors of node $34$, and are assigned to community $C_2$. For the rest three nodes, nodes $17, 25$ and $26$, node $17$ is allocated to community $C_1$ through node $6$, while nodes $25$ and $26$ are allocated to community $C_2$ through node $32$. The final result of our community detection on the karate network suggests that the karate network contains two communities: $C_1=\{1,2,3,4,8,14,18,20,22,5,6,7,11,17\}$ and $C_2=\{9,10,15,16,19,21,23,24,27,28,30,31,33,34,25,26,29,32\}$, which consist of 16 and 18 nodes respectively. This result is consistent with the observation of Zachary.   

For the dolphins network, following exactly the same procedures, we obtained 19 central nodes, and five communities, as shown in figure 4. In contrast, previous literature such as\cite{Lusseau2004} typically splits the dolphins network into four communities. Among them, one community is exactly identical  to our community $C_1$, and other two communities are very close to our communities $C_2$ and $C_3$, except that three nodes, $40, 54$ and $62$, are allocated to community $C_2$ in \cite{Lusseau2004}. In our communities, nodes, $54$ and $62$ are contained in community $C_3$, while node $40$ is contained in community $C_5$. We believe that allocating these two nodes to community $C_3$ is reasonable because node $62$ represents a male dolphin, while the gender of node $54$ is unknown; allocating these two nodes to community $C_3$ makes the number of female dolphins, which is dominant in community $C_2$, grow. The last community in  \cite{Lusseau2004} is technically a combination of our $C_4$ and $C_5$ without node $40$. In a higher level of community structure detected in  \cite{Lusseau2004}, not only $C_4$ and $C_5$, but also $C_1$, $C_2$ and $C_3$ are combined into one bigger community. As a result, the whole network is alternatively split into two communities: $C_1+C_2+C_3$, and $C_4+C_5$, in which node 40 belongs to the former community. To our viewpoint, such a higher level of community structure has a lower value of $F2$ than our community structure, but it is also reasonable. Finally, as for node $40$, it has only two neighbors in the whole network, node $37$ and node $58$. We choose to allocate it to community $C_2$ since its neighbor in $C_2$, node $58$, has a higher value of $LCI$ than node $37$, which is in a different community.

\begin{figure}[H]
\setlength{\abovecaptionskip}{0.cm}
\setlength{\belowcaptionskip}{-0.cm}
\centering
	\includegraphics[width=1\textwidth]{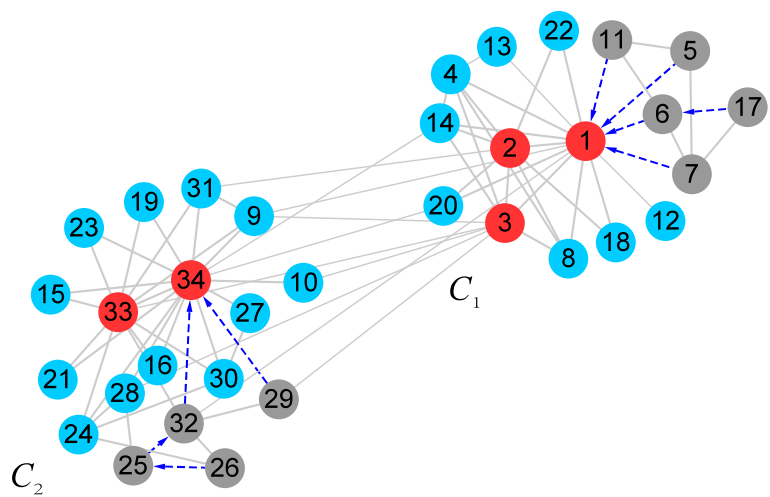}  
    \caption{Community detection on the karate network. Nodes colored by red are central nodes of the network. Two communities of 11 and 14 nodes agglomerate in a greedy community expansion procedure. Blue dashed arrows show the paths through which each of the residual unassigned nodes are allocated to one of the communities. Finally, the extended communities $C_1$ and $C_2$ contain 16 and 18 nodes respectively.}
\label{fig_IIlustration}
\end{figure}

\begin{figure}[H]
\setlength{\abovecaptionskip}{0.cm}
\setlength{\belowcaptionskip}{-0.cm}
\centering
	\includegraphics[width=1\textwidth]{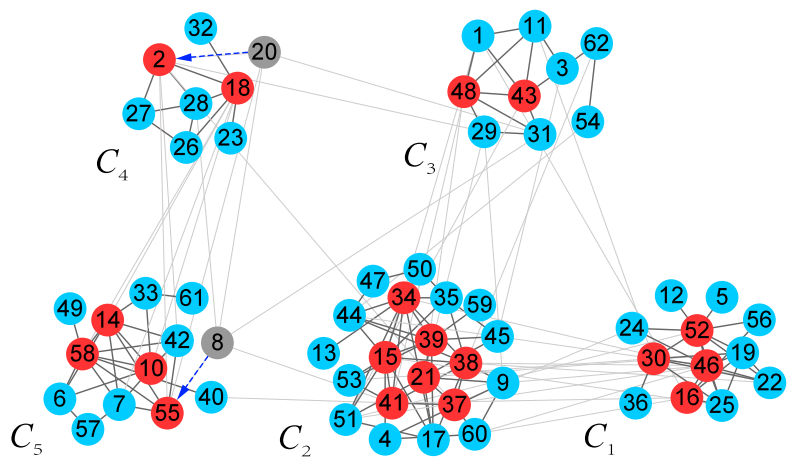}  
    \caption{Community detection on the dolphins network. Following the same procedures as in figure 4, we obtained 19 central nodes colored by red, based on which five communities emerge through community expansion, and blue dashed arrows show the allocation of residual unassigned nodes to communities. 
}
\label{fig_IIlustration}
\end{figure}

On the football network, communities obtained by our algorithm also show minor differences only to the reference community structure reported in \cite{Girvan2002}. Among the total 115 nodes of the network, which distribute in 12 communities, only 8 nodes, $29, 37, 43, 60, 64, 91, 98, 111$, are allocated to different communities in our result than in \cite{Girvan2002}. Specially, in \cite{Girvan2002} nodes $37$, $43$ and $91$ are allocated to the same community as nodes $81$ and $83$. However, there is no connections at all between the former three and latter two nodes. In contrast, our algorithm classifies nodes 81 and 83 as an individual community, while allocates nodes $37$, $43$ and $91$ to other communities that they are actually connected to. Such a result, as we believe, should be more reasonable.


\begin{figure}[H]
\setlength{\abovecaptionskip}{0.cm}
\setlength{\belowcaptionskip}{-0.cm}
\centering
	\includegraphics[width=1\textwidth]{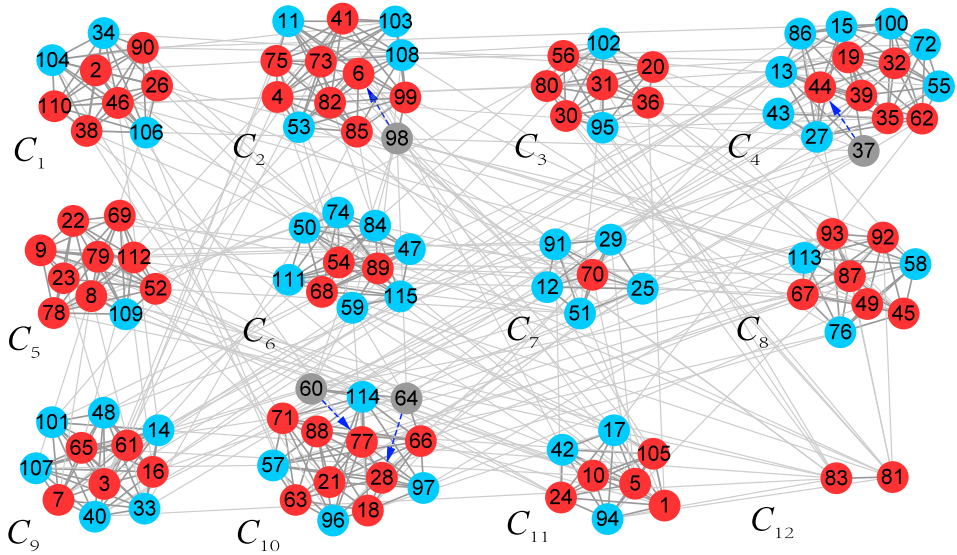}   
    \caption{Community detection on the football network. Following the same procedures as in figures 4 and 5, we finally obtained 12 communities in this network.
}
\label{fig_IIlustration}
\end{figure}

\subsection{the performance of our community detection algorithm}     

We also applied the normalized mutual information ($NMI$) on benchmark networks to evaluate the performance of our community detection algorithm.  

Normalized mutual information ($NMI$) \cite{Kuncheva2004} is an evaluation indicator that measures the performance of a community detection algorithm. On a given network, it compares the community structure detected by a certain algorithm to a standard community structure of the same network, which is used as a reference. We firstly produce a confusion matrix $\textbf{N}$, its element $N_{ij}$ on the $i$th row and $j$th column represents the number of nodes contained in the intersection between the  $i$th community of the reference, and the $j$th community detected by the algorithm to be evaluated. If $N$ stands for the total number of nodes within the network, then $NMI$ is defined as

\begin{equation}
NMI = \frac{{ - 2\sum\limits_{i = 1}^{{C_r}} {\sum\limits_{j = 1}^{{C_f}} {{N_{ij}}N/{N_{i.}}{N_{.j}}} } }}{{\sum\limits_{i = 1}^{{C_r}} {{N_{i.}}\log ({N_{i.}}/N) + \sum\limits_{j = 1}^{{C_f}} {{N_{.j}}\log ({N_{.j}}/N)} } }}
\end{equation}
in which $C_r$ indicates the number of reference communities. $C_f$ denotes the number of communities detected by the algorithm to be evaluated. $N_{i.}$ and $N_{.j}$ stand for the sums of all elements in the $i$th row of $j$th column of $\textbf{N}$ respectively. When the detected communities are exactly identical to the reference, $NMI=1$; on the other hand, when the detected communities are totally independent to the reference, $NMI$ equals to 0. Therefore, $NMI$ reflects the amount of information on the community structure that is correctly extracted by the given algorithm; it is widely used to evaluate the performance of a community detection algorithm.

One kind of classical benchmark networks are LFR networks\cite{Lancichinetti2008}. For LFR networks, degrees of nodes are distributed according to power law with exponent 
$2<\gamma<3$ and the sizes of communities also obey the power law distribution with exponent $1<\beta< 2$. 
Besides, the community size $s$ and node degree $k$ satisfy the constraint $s_{min}>k_{min}$ and $s_{max}>k_{max}$. An important mixing parameter $\mu$ represents the ratio between the external degree of a node with respect to its community and the total degree of the node.  As the value of $\mu$ gets large, the community structure of network becomes ambiguous. 

Now, we can detect the community structure on the LFR network with our greedy algorithm employing $LCI$ and $F2$, and then calculate the $NMI$ to evaluate the performance of the algorithm. Figure 6(a) shows the $NMI$ varying with the increase of the mixing parameter $\mu$ on the LFR networks; network parameters are as following: total number of nodes $N=500, 1000, 2000$ and $5000$, average degree  $\langle k\rangle=20$, $\gamma=2.5$, $\beta=1.5$. When $\mu$ is small, the $NMI$ is close to $1$. With the increase of $\mu$, the community structure becomes more and more ambiguous and hard to detect, the $NMI$ gradually decreases. Figure 6(a) shows that our method performs better on larger networks. The reason is, with the same average degree  $\langle k\rangle$, larger networks usually contain more communities. Under the same value of $\mu$, the external links from a node within a certain community tend to be distributed to different other communities. Relatively, the internal links from the same node will appear more concentrated, which makes the community structure more distinct and easy to detect. In figure 6(b), we compare the $NMI$s with $N=500$ for community detection algorithms based on different modularity functions, $Q$, $F$, $M$, $R$,and $F2$. Obviously, the algorithm based on $F2$ performs the best since it shows the highest $NMI$.

$NMI$ can also be calculated for the real-world networks. Using the community structures suggested in \cite{Zachary1977,Girvan2002,Lusseau2004} as references, our communities detected for the karate network, the dolphins network and the football network respectively show the following $NMI$s: $1$ for karate, $0.8904$ for dolphins, and $0.9429$ for football (see table 1). Obviously communities detected by our method are all highly consistent with the reference communities suggested in previous literature.

\begin{figure}[H]
\setlength{\abovecaptionskip}{0.cm}
\setlength{\belowcaptionskip}{-0.cm}
\centering
	\includegraphics[width=1\textwidth]{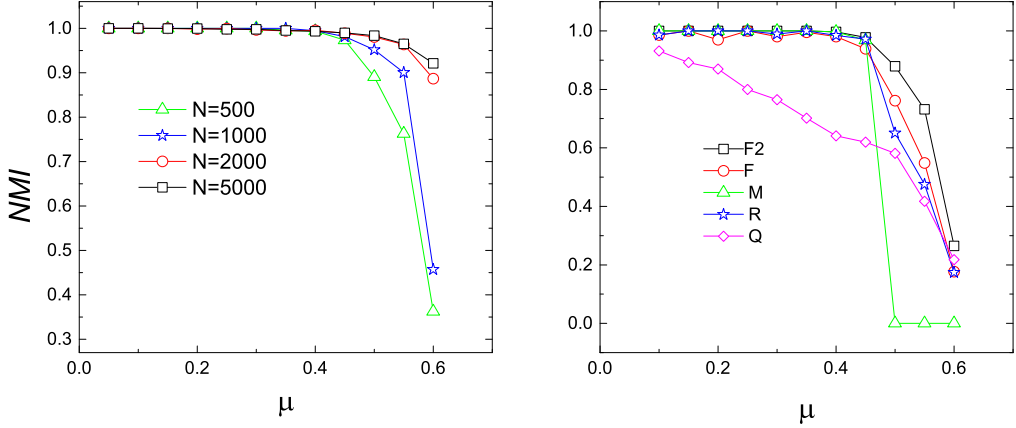}
    \caption{$NMI$s of different algorithms on the LFR networks. Parameters of the LFR networks are set as: average degree $\langle k\rangle=20$,  $\gamma=2.5$, $\beta=1.5$, and varying network size $N$. (a) $NMI$s for our community detection algorithm based on $LCI$ and $F2$, with $N=500, 1000, 2000$ and $5000$. (b) $NMI$s for different community detection algorithms based on different modularity functions, $Q$, $M$, $F$, $R$ and $F2$, with $N=500$. 
}
\label{fig_IIlustration}
\end{figure}

\section{Conclusion}

In this paper, we proposed a new community detection method to identify local communities within complex networks. We firstly suggested a new local centrality indicator ($LCI$) to extract local important nodes that are well distinguished from their neighbors. Compared with the global maximal degree and local maximal degree indicators\cite{Chen2013}, our $LCI$ extracts central nodes that can be directly connected but have different degrees. Then we proposed a new local modularity function $F2$.  $F2$ can overcome certain problems of other modularity functions such as the resolution limit problem. Both theoretical deductions and numerical simulations suggest that $F2$ identifies communities, as well as the widely-used modularity functions $Q, M, F, R$. In certain cases, $F2$ performs even better than $Q, M, F$ and $R$. Incorporated to a greedy algorithm, $LCI$ and $F2$ identify communities within a complex network  automatically. On both the real-world networks and the computer simulated benchmark networks, our method shows high performance of identifying community structures. 

Through proper revisions, our modularity function $F2$ can be easily extended to directed or weighted networks, or to deal with communities that have overlaps. With the development of big data, computation efficiency has become a coming demand. On the other hand, dynamical networks and dynamical community detection has also become a worthwhile research area. In interdisciplinary area, community detection has been applied to find the pathways between given diseases and drugs\cite{Pham2019}, and to reveal the role of each part of a layered neural network by analyzing communities extracting from the trained network\cite{Watanabe2019}. Community detection is also employed to mine user opinions from social networks\cite{Li2019}. Besides, modularity functions can be used to assess the training results for neural networks\cite{Watanabe2018}. Our study on the modularity function $F2$ may hopefully lead to further researches that might be worth pursuing.

\section*{Acknowledgements}
\addcontentsline{toc}{section}{Acknowledgements}
This work was supported by the National Natural Science Foundation of China with Grant No. 61673085, 61703074, the "Thousand Talents Program" of Sichuan Province with Grant No.17QR003 , P.R. China and the fund of the Education Department of Sichuan Province with Grant No.13ZA0168.

\section*{Appendix. Characteristics of the modularity function $F2$}

(1) A random network can't be partitioned to two separated communities. When the whole network is considered as one community, $F2_{single}=1$. When it is divided into any two parts, which contain $n_1$ and $n_2$ nodes respectively, the modularity function $F2$ can be calculated as 


\begin{equation}
\begin{array}{l}
F2_{merge} = {\left[ {\frac{{{n_1}({n_1} - 1)p}}{{{n_1}({n_1} - 1)p + {n_1}{n_2}p}}} \right]^2} + {\left[ {\frac{{{n_2}({n_2} - 1)p}}{{{n_2}({n_2} - 1)p + {n_1}{n_2}p}}} \right]^2}\\
{\kern 1pt} {\kern 1pt} {\kern 1pt} {\kern 1pt} {\kern 1pt} {\kern 1pt} {\kern 1pt} {\kern 1pt} {\kern 1pt} {\kern 1pt} {\kern 1pt} {\kern 1pt} {\kern 1pt} {\kern 1pt} {\kern 1pt} {\kern 1pt} {\kern 1pt} {\kern 1pt}  = {\left( {\frac{{{n_1} - 1}}{{{n_1} + {n_2} - 1}}} \right)^2} + {\left( {\frac{{{n_2} - 1}}{{{n_2} + {n_1} - 1}}} \right)^2}\\
{\kern 1pt} {\kern 1pt} {\kern 1pt} {\kern 1pt} {\kern 1pt} {\kern 1pt} {\kern 1pt} {\kern 1pt} {\kern 1pt} {\kern 1pt} {\kern 1pt} {\kern 1pt} {\kern 1pt} {\kern 1pt} {\kern 1pt} {\kern 1pt}  = \frac{{n_1^2 + n_2^2 - 2{n_1} - 2{n_2} + 2}}{{{{({n_1} + {n_2} - 1)}^2}}}\\
{\kern 1pt} {\kern 1pt} {\kern 1pt} {\kern 1pt} {\kern 1pt} {\kern 1pt} {\kern 1pt} {\kern 1pt} {\kern 1pt} {\kern 1pt} {\kern 1pt} {\kern 1pt} {\kern 1pt} {\kern 1pt} {\kern 1pt}  = 1 - \frac{{2{n_1}{n_2} - 1}}{{{{({n_1} + {n_2} - 1)}^2}}}\\
{\kern 1pt} {\kern 1pt} {\kern 1pt} {\kern 1pt} {\kern 1pt} {\kern 1pt} {\kern 1pt} {\kern 1pt} {\kern 1pt} {\kern 1pt} {\kern 1pt} {\kern 1pt} {\kern 1pt} {\kern 1pt}  < 1
\end{array}
\end{equation}

(2) A complete network can't be partitioned to two communities, Obviously, the $F2$ for a complete network identified as a single community is $F2_{single}=1$.If
the network is divided to any two communities which include $n_1$ and $n_2$ nodes respectively, the $F2$ can be calculated as 


\begin{equation}
\begin{array}{l}
F2_{merge} = {\left[ {\frac{{{n_1}({n_1} - 1)}}{{{n_1}({n_1} - 1) + {n_1}{n_2}}}} \right]^2} + {\left[ {\frac{{{n_2}({n_2} - 1)}}{{{n_2}({n_2} - 1) + {n_1}{n_2}}}} \right]^2}\\
{\kern 1pt} {\kern 1pt} {\kern 1pt} {\kern 1pt} {\kern 1pt} {\kern 1pt} {\kern 1pt} {\kern 1pt} {\kern 1pt} {\kern 1pt} {\kern 1pt} {\kern 1pt} {\kern 1pt} {\kern 1pt} {\kern 1pt} {\kern 1pt} {\kern 1pt} {\kern 1pt} 
 = {\left( {\frac{{{n_1} - 1}}{{{n_1} + {n_2} - 1}}} \right)^2} + {\left( {\frac{{{n_2} - 1}}{{{n_2} + {n_1} - 1}}} \right)^2}\\
{\kern 1pt} {\kern 1pt} {\kern 1pt} {\kern 1pt} {\kern 1pt} {\kern 1pt} {\kern 1pt} {\kern 1pt} {\kern 1pt} {\kern 1pt} {\kern 1pt} {\kern 1pt} {\kern 1pt} {\kern 1pt} {\kern 1pt} {\kern 1pt}  
 < 1
\end{array}
\end{equation}

(3) For a ring network constructed by $l$ $p$-cliques, similar to the figure 2(a), the network can't two or more cliques into one community. If each community corresponds a single clique, the $F2$ is,

\begin{equation}
F2_{single} = l \times {\left[ {\frac{{p(p - 1)}}{{p(p - 1) + 2}}} \right]^2}
\end{equation}

If we merge $h$ $p$-cliques into one community, the value of $F2$ becomes,

\begin{equation}
\begin{array}{l}
F2_{merge} = \frac{N}{h}{\left[ {\frac{{p(p - 1)h + 2(h - 1)}}{{p(p - 1)h + 2(h - 1) + 2}}} \right]^2}\\
{\kern 1pt} {\kern 1pt} {\kern 1pt} {\kern 1pt} {\kern 1pt} {\kern 1pt} {\kern 1pt} {\kern 1pt} {\kern 1pt} {\kern 1pt} {\kern 1pt} {\kern 1pt} {\kern 1pt} {\kern 1pt} {\kern 1pt} {\kern 1pt} {\kern 1pt} {\kern 1pt}  = \frac{N}{h}{\left[ {\frac{{p(p - 1)h + 2(h - 1)}}{{p(p - 1)h + 2h}}} \right]^2}\\
{\kern 1pt} {\kern 1pt} {\kern 1pt} {\kern 1pt} {\kern 1pt} {\kern 1pt} {\kern 1pt} {\kern 1pt} {\kern 1pt} {\kern 1pt} {\kern 1pt} {\kern 1pt} {\kern 1pt} {\kern 1pt} {\kern 1pt} {\kern 1pt}  = N{\left[ {\frac{{\frac{{p(p - 1)h + 2h - 2}}{{{h^3}}}}}{{p(p - 1) + 2}}} \right]^2}\\
{\kern 1pt} {\kern 1pt} {\kern 1pt} {\kern 1pt} {\kern 1pt} {\kern 1pt} {\kern 1pt} {\kern 1pt} {\kern 1pt} {\kern 1pt} {\kern 1pt} {\kern 1pt} {\kern 1pt} {\kern 1pt} {\kern 1pt}  \le N{\left[ {\frac{{\frac{{p(p - 1) + 1}}{4}}}{{p(p - 1) + 2}}} \right]^2}{\kern 1pt} {\kern 1pt} {\kern 1pt} {\kern 1pt} (h \ge 2)\\
{\kern 1pt} {\kern 1pt} {\kern 1pt} {\kern 1pt} {\kern 1pt} {\kern 1pt} {\kern 1pt} {\kern 1pt} {\kern 1pt} {\kern 1pt} {\kern 1pt} {\kern 1pt} {\kern 1pt} {\kern 1pt}  = N{\left[ {\frac{{p(p - 1) + \frac{{1 - 3p(p - 1)}}{4}}}{{p(p - 1) + 2}}} \right]^2}{\kern 1pt} {\kern 1pt} {\kern 1pt} \\
{\kern 1pt} {\kern 1pt} {\kern 1pt} {\kern 1pt} {\kern 1pt} {\kern 1pt} {\kern 1pt} {\kern 1pt} {\kern 1pt} {\kern 1pt} {\kern 1pt} {\kern 1pt} {\kern 1pt} {\kern 1pt}  < N{\left[ {\frac{{p(p - 1)}}{{p(p - 1) + 2}}} \right]^2}{\kern 1pt} {\kern 1pt} {\kern 1pt} (p \ge 3)
\end{array}
\end{equation}

(4) Communities with different scales can be identified by $F2$. As shown in figure 2(b), when two $q$-cliques are considered as two separated communities, the value of $F2$ is,

\begin{equation}
\begin{array}{l}
F2_{single} = {\left[ {\frac{{{p}(p - 1)}}{{p(p - 1) + 1}}} \right]^2}
 +{\left[ {\frac{{{p}(p - 1)}}{{p(p - 1) + 3}}} \right]^2}
 + 2 \times {\left[ {\frac{{{q}(q - 1)}}{{q(q - 1) + 2}}} \right]^2}\\
{\kern 1pt} {\kern 1pt} {\kern 1pt} {\kern 1pt} {\kern 1pt} {\kern 1pt} {\kern 1pt} {\kern 1pt} {\kern 1pt} {\kern 1pt} {\kern 1pt} {\kern 1pt} {\kern 1pt} {\kern 1pt} {\kern 1pt} {\kern 1pt} {\kern 1pt}  = 
{\left[{\frac{{{p}(p - 1)}}{{p(p - 1) + 1}}} \right]^2}
 +{\left[ {\frac{{{p}(p - 1)}}{{p(p - 1) + 3}}} \right]^2}
 + \frac{{{{[q(q - 1)]}^2} + [q(q - 1)] \times [q(q - 1)]}}{{{{[q(q - 1) + 2]}^2}}}
\end{array}
\end{equation}

However, if two $q$-cliques are merged into one community, the network has a $F2$ value

\begin{equation}
\begin{array}{l}
F2_{merge} =  {\left[ {\frac{{{p}(p - 1)}}{{p(p - 1) + 1}}} \right]^2}
 +{\left[ {\frac{{{p}(p - 1)}}{{p(p - 1) + 3}}} \right]^2}
 + {\left[ {\frac{{2{q}({q} - 1) + 2}}{{2{q}({q} - 1) + 4}}} \right]^2}\\
{\kern 1pt} {\kern 1pt} {\kern 1pt} {\kern 1pt} {\kern 1pt} {\kern 1pt} {\kern 1pt} {\kern 1pt} {\kern 1pt} {\kern 1pt} {\kern 1pt} {\kern 1pt} {\kern 1pt} {\kern 1pt} {\kern 1pt} {\kern 1pt} {\kern 1pt} {\kern 1pt}  = 
 {\left[ {\frac{{{p}(p - 1)}}{{p(p - 1) + 1}}} \right]^2}
 +{\left[ {\frac{{{p}(p - 1)}}{{p(p - 1) + 3}}} \right]^2}
 + {\left[ {\frac{{{q}({q} - 1) + 1}}{{{q}({q} - 1) + 2}}} \right]^2}\\
{\kern 1pt} {\kern 1pt} {\kern 1pt} {\kern 1pt} {\kern 1pt} {\kern 1pt} {\kern 1pt} {\kern 1pt} {\kern 1pt} {\kern 1pt} {\kern 1pt} {\kern 1pt} {\kern 1pt} {\kern 1pt} {\kern 1pt} {\kern 1pt} {\kern 1pt} {\kern 1pt}  =
 {\left[ {\frac{{{p}(p - 1)}}{{p(p - 1) + 1}}} \right]^2}
 +{\left[ {\frac{{{p}(p - 1)}}{{p(p - 1) + 3}}} \right]^2}
 + \frac{{{{[{q}({q} - 1)]}^2} + 2{q}({q} - 1) + 1}}{{{{[{q}({q} - 1) + 2]}^2}}}\\
{\kern 1pt} {\kern 1pt} {\kern 1pt} {\kern 1pt} {\kern 1pt} {\kern 1pt} {\kern 1pt} {\kern 1pt} {\kern 1pt} {\kern 1pt} {\kern 1pt} {\kern 1pt} {\kern 1pt} {\kern 1pt} {\kern 1pt} {\kern 1pt} {\kern 1pt}  < 
 {\left[ {\frac{{{p}(p - 1)}}{{p(p - 1) + 1}}} \right]^2}
 +{\left[ {\frac{{{p}(p - 1)}}{{p(p - 1) + 3}}} \right]^2}
 + \frac{{{{[{q}(q - 1)]}^2} + [q(q - 1)] \times [q(q - 1)]}}{{{{[q(q - 1) + 2]}^2}}}{\kern 1pt} {\kern 1pt} {\kern 1pt} {\kern 1pt} {\kern 1pt} {\kern 1pt} ({q} \ge 3)
\end{array}
\end{equation}

\end{document}